\documentclass[twoside]{article}
\usepackage{fleqn,espcrc2,epsf}

\usepackage{graphicx}
\usepackage[figuresright]{rotating}

\newcommand{\AmS}{{\protect\the\textfont2
  A\kern-.1667em\lower.5ex\hbox{M}\kern-.125emS}}

\title{Heavy-light meson spectrum with and without NRQCD}

\author{Randy Lewis\address{Department of Physics, University of Regina,
        Regina, SK, S4S 0A2, Canada}
    and R. M. Woloshyn\address{TRIUMF, 4004 Wesbrook Mall, Vancouver, BC,
        V6T 2A3, Canada}}
       
\begin{document}

\begin{abstract}
Results for the spectrum of S and P-wave charmed mesons are 
obtained in the quenched approximation from a tadpole-improved 
anisotropic gauge field action and a D234 quark action. 
This is compared to the spectrum obtained from an NRQCD charm 
quark and a D234 light antiquark.  NRQCD results for bottom mesons
are also discussed.
\vspace{1pc}
\end{abstract}

\maketitle

\section{MOTIVATION}

If the charm quark is sufficiently heavy, then lattice NRQCD is an efficient
computational method for obtaining the spectrum of mesons with
a single charm quark.  Recent studies of the S and P-wave meson
spectrum have determined the contributions from first, second and third 
order in the inverse charm mass expansion.\cite{ourSwaves,ourPwaves}
Second and third order contributions were found to be statistically 
insignificant for the P-wave mesons, but the S-wave spin splittings 
($D^*-D$ and $D_s^*-D_s$) acquire substantial corrections at both
second and third order.  It is difficult to make a firm statement about
convergence of the expansion from these findings alone.

In the present work, the existing results are compared to a new study,
where the NRQCD charm quark is replaced by a relativistic charm quark
using a D234 fermion action.\cite{D234}
The new simulations use the same set of gauge field configurations and
the same light antiquark propagators as were used in Ref.~\cite{ourPwaves},
so the only differences are the heavy quark action (NRQCD versus D234)
and the operators used to create or destroy the various meson states.

The S-wave masses have been discussed many times in the 
literature (see Refs.~\cite{ourSwaves,ourPwaves,Hein} and references therein), 
but the P-waves have only recently been
extracted from lattice data: the limit of an infinitely heavy quark
was studied by Michael and Peisa\cite{MicP}, a subset of the P-wave
charmed masses were found by Boyle using the clover action (he also
extrapolated to the bottom region)\cite{Boyle}, and two groups have
used NRQCD to study the charmed and bottom meson 
spectra\cite{ourPwaves,Hein,lat99,AliKhan}.

\section{LATTICE CHOICES AND METHOD}

The gauge fields of Ref.~\cite{ourPwaves}, also used for the present study,
are on $10^3\times30$ anisotropic lattices with $(a_s/a_t)_{\rm bare} = 2$.
The gauge action is classically improved by
including rectangular plaquettes, and tadpole improved using the mean link
in Landau gauge.  The coupling is fixed at $\beta=2.1$, and the renormalized
anisotropy is found to be $a_s/a_t = 1.96(2)$.

The light antiquark fields are also taken from Ref.~\cite{ourPwaves}.
They are derived from a D234 action\cite{D234} which removes leading and 
next-to-leading classical errors and is tadpole improved.  Two hopping 
parameters ($\kappa = 0.23$ and 0.24) are used, 
corresponding to ``$m_\pi/m_\rho$'' = 0.815(3) and 
0.517(8).  The physical $\rho$ meson mass leads to $a_t = 0.1075(23)$ fm,
and the physical kaon mass implies $\kappa_{\rm strange} = 0.2356(3)$.
Dirichlet time boundaries are used for the light antiquark.

In Ref.~\cite{ourPwaves} the heavy quark was described by NRQCD, and new
results are presented here where a heavy D234 quark is used with 
$\kappa=0.182$, which is in the vicinity of charm.  By interpolating the NRQCD
results to this same heavy quark mass, a direct comparison of NRQCD and D234
results can be made.

\begin{table}[htb]
\caption{Heavy-light creation operators.  $\Delta_i$ is a spatial lattice
         derivative.}\label{ops}
\begin{tabular}{lll}
\hline
                  & NRQCD & D234 \\
   \cline{2-3}
   ${}^{2S+1}L_J$ & $\Omega(\vec x)$ & $\Omega(x)$ \\
   \hline
   ${}^1S_0$ & $(~0,~I~)$ & $\gamma_5$ \\
   ${}^3S_1$ & $(~0,~\sigma_i~)$ & $\gamma_i$ \\
   ${}^1P_1$ & $(~0,~\Delta_i~)$ & $\sigma_{ij}$ \\
   ${}^3P_0$ & $(~0,~\sum_i\Delta_i\sigma_i~)$ & $I$ \\
   ${}^3P_1$ & $(~0,~\Delta_i\sigma_j-\Delta_j\sigma_i~)$ &$\gamma_i\gamma_5$\\
   ${}^3P_2$ & $(~0,~\Delta_i\sigma_i-\Delta_j\sigma_j~)$ &
               $\gamma_i\Delta_i-\gamma_j\Delta_j$ \\
             & or $(~0,~\Delta_i\sigma_j+\Delta_j\sigma_i~)$ \\
\hline
\end{tabular}
\end{table}
With and without NRQCD, heavy-light meson masses come from 2000 quenched 
configurations, and
statistical uncertainties are obtained from 5000 bootstrap ensembles.
The heavy-light meson creation operators are
\begin{equation}
\sum_{\vec x}Q^\dagger(\vec x)\Omega(\vec x)
   [1+c_s\Delta^{(2)}(\vec x)]^{n_s}q(\vec x)
\end{equation}
for NRQCD, and
\begin{equation}
\bar\psi(x)\Omega(x)\left[1+c_sF(x)\right]^{n_s}\psi(x)
\end{equation}
for D234, where
\begin{equation}
F(x) = \sum_{i=1}^3\left[U_i(x)\delta_{x,y-\hat{i}}
       +U_i^\dagger(x-\hat{i})\delta_{x,y+\hat{i}}\right]
\end{equation}
and $\Delta^{(2)}$ is the lattice Laplacian.

The quantum numbers are dictated by $\Omega$ as presented in Table~\ref{ops}.
At the source, the parameters $(c_s,n_s)$ are set to (0.15,10) for NRQCD 
and (1,15) for D234.  The sink operator is local in all simulations.

\section{S-WAVE HEAVY-LIGHT MASSES}

\begin{figure}[thb]
\epsfxsize=400pt \epsfbox[80 400 980 730]{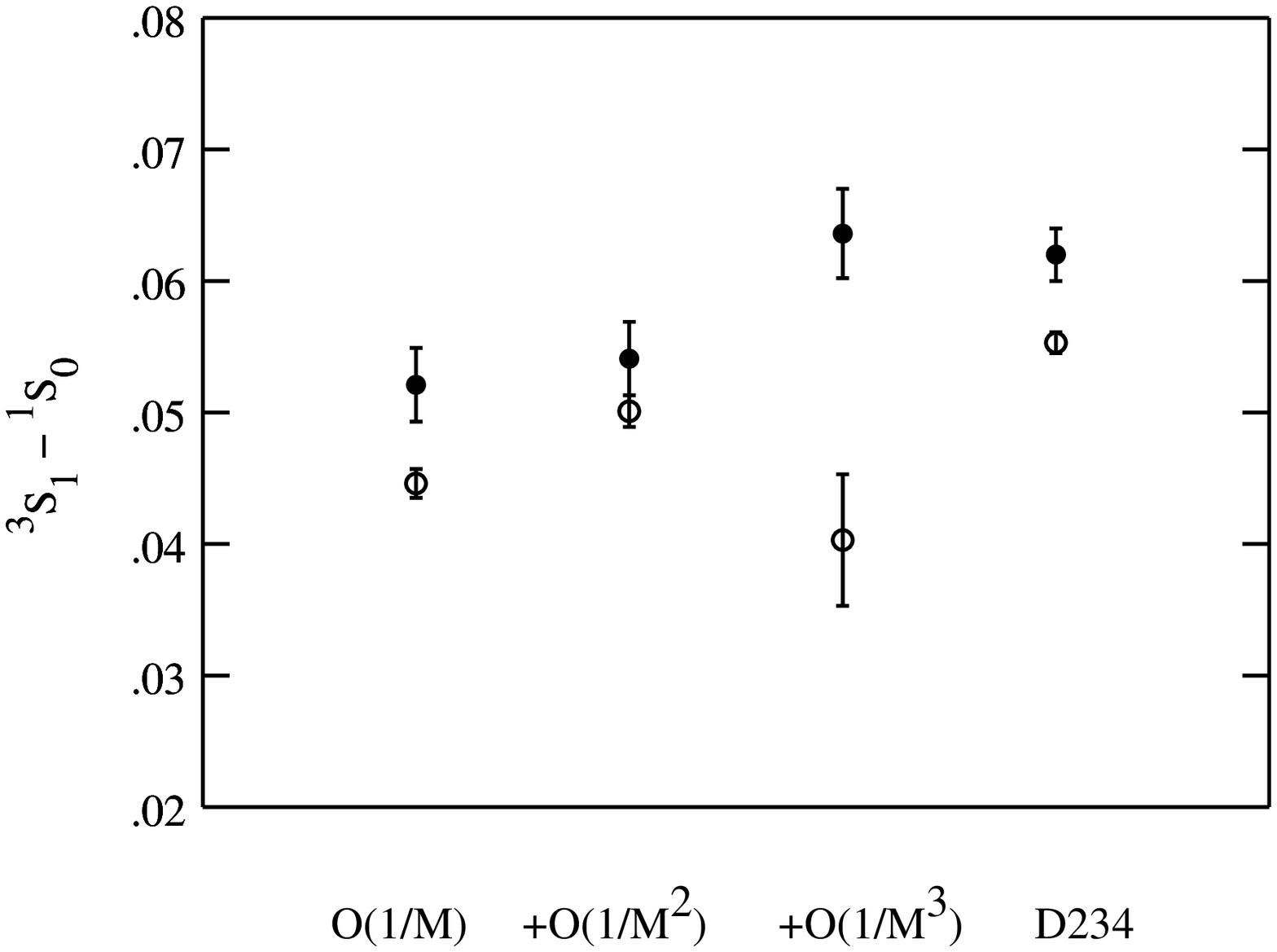}
\caption{Comparison of D234 and NRQCD for the S-wave spin splitting, in
         lattice units.
         Solid(open) symbols use $\kappa=0.23(0.24)$ for the light antiquark.
         \vspace{-5mm}}
\label{fig:scompare}
\end{figure}

The S-wave hyperfine splitting for D234 is shown in Fig.~\ref{fig:scompare}
in lattice units, for the two available light antiquark hopping parameters.
Also shown are the NRQCD results from Ref.~\cite{ourPwaves} after interpolation
of the ${}^1S_0$ kinetic mass to match the D234 value of the ${}^1S_0$ meson
mass.  This interpolation
was done independently at each order in the NRQCD expansion, so the
${}^1S_0$ mass maintains a common value in each case.  
From Fig.~\ref{fig:scompare}, lowest order NRQCD
is seen to give smaller hyperfine splittings than D234.  The addition of
$O(1/M^2)$ terms tends to increase the NRQCD results, but $O(1/M^3)$
contributions are larger in magnitude than the preceding order so the
status of the NRQCD expansion for this observable is unclear.

\begin{figure}[htb]
\epsfxsize=400pt \epsfbox[80 400 980 730]{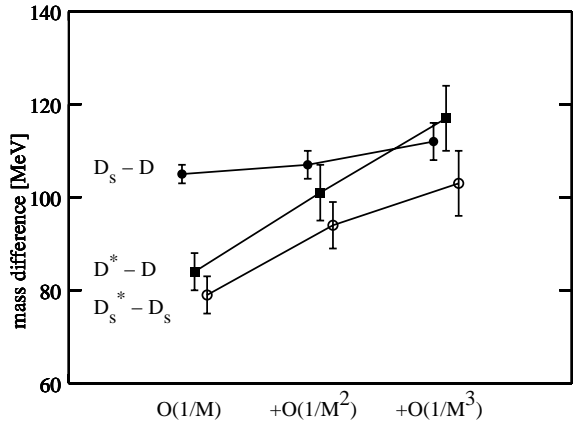}
\caption{NRQCD S-wave mass splittings in physical units.\vspace{-2mm}}
\label{fig:sdiff}
\end{figure}

The NRQCD results are shown in physical units in Fig.~\ref{fig:sdiff}.
Whereas Fig.~\ref{fig:scompare} compared the different NRQCD orders by
choosing a common physical ${}^1S_0$ mass, Fig.~\ref{fig:sdiff} shows
the NRQCD expansion when the bare charm quark mass is held fixed.
Neither plot gives a compelling defense for a convergent $1/M_{\rm charm}$ 
expansion,
but convergence is not disproven by such plots either.  The $D_s-D$ splitting
receives insignificant corrections from second and third orders, but
corrections to the spin splittings are significant.  Experimentally
the $D_s-D^+$ splitting is 104 MeV, and therefore agrees with the NRQCD
determination, but the spin splittings are $D^{*+}-D^+ = 141$ MeV and
$D_s^*-D_s = 144$ MeV.  With or without NRQCD, the quenched prediction
is smaller than experiment.

Ref.~\cite{ourPwaves} also provides results for the bottom mesons, and
finds that the S-wave bottom masses are completely dominated by leading order 
in NRQCD so convergence is not disputed.  As for charm, the spin splittings 
are smaller than experiment.

\section{P-WAVE HEAVY-LIGHT MASSES}

\begin{figure}[thb]
\epsfxsize=400pt \epsfbox[80 400 980 730]{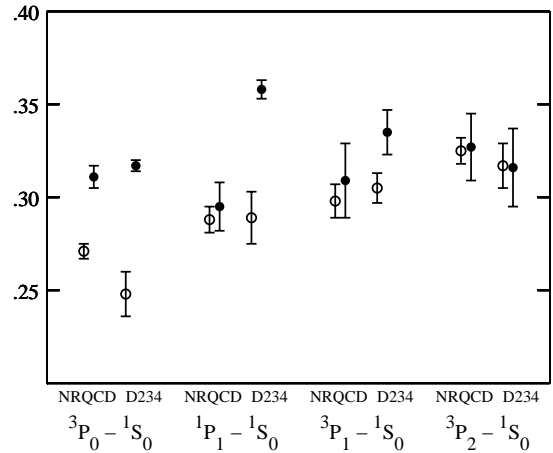}
\caption{Comparison of D234 and NRQCD for the S-P splittings, in lattice
         units.
         Solid(open) symbols use $\kappa = 0.23(0.24)$ for the light
         antiquark.\vspace{-4mm}}
\label{fig:pcompare}
\end{figure}

In Ref.~\cite{ourPwaves}, nonleading corrections to P-wave
heavy-light masses were found to be insignificant for both bottom and
charm mesons.  This leads one to expect agreement between D234 and
NRQCD determinations of these P-wave masses.
Fig.~\ref{fig:pcompare} shows all four of the P-wave masses in lattice units
for both of the available light antiquark hopping parameters.  D234
results are new, and the NRQCD results are interpolations of the data
from Ref.~\cite{ourPwaves} so that the ${}^1S_0$ kinetic mass matches the
D234 ${}^1S_0$ mass.  Only statistical uncertainties are shown.

According to Fig.~\ref{fig:pcompare}, the P-wave masses do not depend upon
whether the charm quark is described by NRQCD or D234.
One rather dramatic exception seems occur for the ${}^1P_1$ with the 
heavier antiquark, but this systematic error is understood: the D234
calculation has no true plateau in this case.
The plateau-finding algorithm of Ref.~\cite{ourPwaves}
has been carried over to the present work, and it chooses the best plateau
by maximizing the quality factor,
\begin{equation}
Q \equiv \frac{\Gamma(N/2-1,\chi^2/2)}{\Gamma(N/2-1,0)},
\end{equation}
where 
\begin{equation}
\Gamma(a,x) = \int_x^\infty{\rm d}t\,t^{a-1}\exp(-t)
\end{equation}
and $N$ is the number of timesteps in the proposed plateau region.
The D234 ${}^1P_1$ meson becomes quite noisy at early Euclidean times
such that no clear plateau is evident,
so the method of maximum $Q$ chooses a ``plateau'' which begins too near the
source.  This leads to the erroneously large mass shown in 
Fig.~\ref{fig:pcompare}.  Except for this one correlator, 
Fig.~\ref{fig:pcompare} presents a clear agreement 
between results with and without NRQCD.

\begin{figure}[thb]
\epsfxsize=400pt \epsfbox[80 70 980 750]{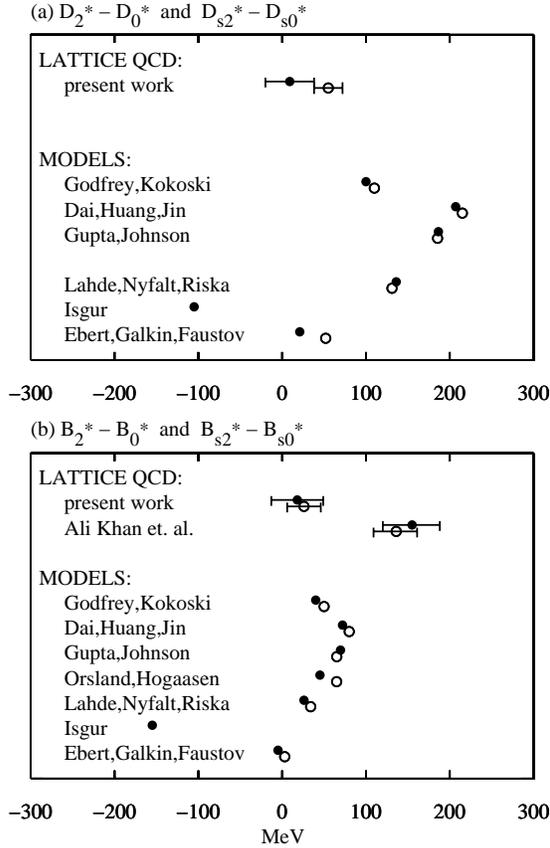}
\caption{Splittings between the ${}^3P_2$ and ${}^3P_0$ masses for the
         $D^{**}$ and $B^{**}$ systems.  Open(solid) symbols involve an
         s(u,d) quark.}
\label{fig:models}
\end{figure}

\begin{figure}[thb]
\epsfxsize=400pt \epsfbox[80 70 980 750]{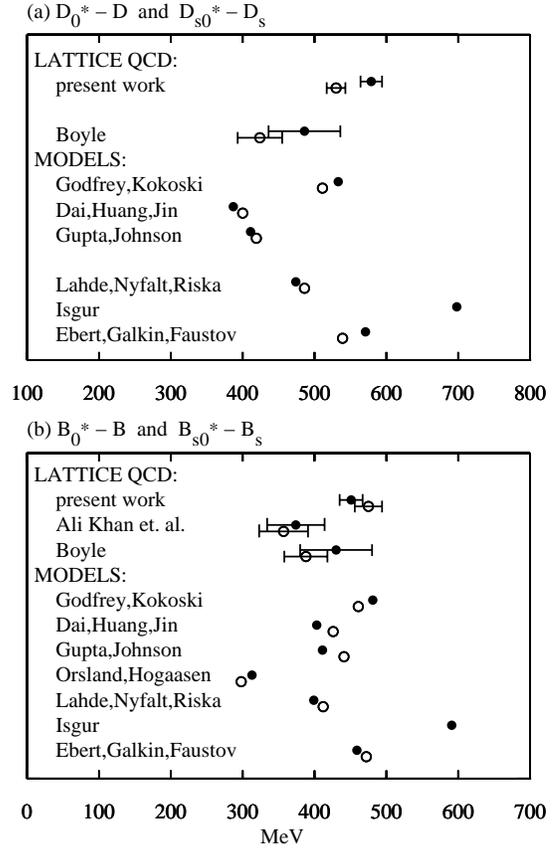}
\caption{S-P splittings for heavy-light mesons.
         Open(solid) symbols involve an s(u,d) quark.}
\label{fig:modelP0}
\end{figure}

The ${}^3P_0$ and ${}^3P_2$ masses for charm and bottom are displayed
in Figs.~\ref{fig:models} and \ref{fig:modelP0} from the NRQCD results
of Ref.~\cite{ourPwaves}, along with NRQCD results from Ref.~\cite{AliKhan},
the clover results of Ref.~\cite{Boyle}, and a variety of model
calculations\cite{models}.
A disagreement between the two lattice NRQCD calculations for $B_2^*-B_0^*$
is evident; Ref.~\cite{ourPwaves} notes that no choice of plateau
region allows those data to attain the large splitting of
Ref.~\cite{AliKhan}.  The lattice results of Ref.~\cite{Hein} cannot
resolve the discrepancy.
The nonlattice models are more easily distinguished from one another
for charm than for bottom, and the lattice result clearly
favours a small $D_2^*-D_0^*$ splitting (substantially less than 100 MeV).

\vspace{3mm}
\noindent
{\bf Acknowledgements}

This work was supported in part by the Natural Sciences
and Engineering Research Council of Canada.

\end{document}